\documentclass[aps]{article}
\usepackage{geometry}              
\usepackage{graphicx}
\usepackage{subfig}
\usepackage{amssymb}
\usepackage{epstopdf}
\usepackage{amsmath}
\usepackage{fixmath}
\usepackage{MnSymbol}
\usepackage{amsfonts}
\usepackage{bm}
\DeclareMathAlphabet{\mathpzc}{OT1}{pzc}{m}{it}

\title{Hidden Invariants in Rheology}
\author{Clifford Chafin\\
\small{Department of Physics, North Carolina State University, Raleigh, NC 27695} \thanks{cechafin@ncsu.edu}}

\begin{document}
\maketitle
\begin{abstract}
This article will use arguments from the deformation driven component of mixing, especially important for microfluidics, to show that the standard invariant based approaches to rheology are lacking.  It is shown that the deviator, $D_{ij}$, after the process of symmetrization, loses microscopically determined information that distinguish rotation from shear and extension in a unique fashion.  We recover this information through an analysis of the discrete processes that must underlie deformation driven mixing in highly correlated fluids.  Without this we show there is no hope of ever deriving adequate general material parameters for rheology from microscopic dynamics.  There is an unambiguous microscopic notion of the rotation rate for every parcel and we derive a general class of invariant rheological theories from it.  We discuss some implications of higher order flows on solutions and suspensions including possibilities for driving and stabilization of nonuniform distributions using hydrodynamic forces alone.  
\end{abstract}

Hydrodynamics, in the case of gases, can be thought of as the lowest order corrections to the stationary solutions to the Maxwell-Boltzmann equations.  The variation of the velocity flow field on the scale of the mean free path is assumed very small compared to the thermal velocity.  Despite the relatively slow speeds of these flows compared to the particles, the collision frustrated diffusive processes generally transport mass, energy and mix much less quickly than these long range convective flows.  In granular materials, jamming obstructs internal motions unless the system is sufficiently agitated.  Liquids have similarly high densities so the local obstructions are important but thermal motions are strong enough to overcome transient bonding and obstructions to allow rearrangements easily enough that static shear stress cannot exist.  
In the case of complex fluids, large and often deformable molecules can have slow relaxation times so that the flow itself alters the orientation, extension, entanglement and local variability in these quantities.  This begs the question of how to relate the microscopic properties to the larger scale hydrodynamics, rheology, of the liquid.  

One way to attack this is with a local set of statistical variables that describe the microscopic configurations and a constitutive law that the relates them to the local flow and local composition of the fluid.  This must ultimately yield the internal stresses.  Such a theory must  incorporate a relaxation rate to give a unique relaxed local configuration for each given local flow.  This seems to provide the most direct connection between kinetics and hydrodynamics.  However, the microscopic details of liquids remain a difficult problem and it is conceivable that we are lacking some fundamental aspect of their nature in our models.  Additionally, in the case of polymers, the local arrangements can be extraordinarily complex and a simple set of variables sufficient for their description is not obvious from microscopic simulations.  

The modern approach for rheology stems from the theory of invariants \cite{Oldroyd} \cite{Truesdell}.  This gives a way to look at a symmetry restricted low energy set of corrections that then depend on a, hopefully, small set of parameters to fit the data.  This attitude is embodied in the modern approach to effective field theory.  In practice, such models have had mixed success and the number of parameters can grow rapidly.  One has to wonder if we have a truly general description and so many parameters are needed or if our model has left something out.  Before we address this let us consider another aspect of fluids: mixing.  

Mixing of constituents, impurities and heat in liquids is generally thought of in terms of diffusion or turbulence.  Turbulence gives a kind of folding map in the manner of the Smale horseshoe \cite{Smale}.  This brings different regions in close proximity so that the slower diffusive process can microscopically mix them.  The dynamics of granular materials are very different than liquids, most notably, the latter cannot withstand static shear forces.  However, they both are restricted to local obstructions and correlations in their motion.  The constituent particles have well defined localization relative to their neighbors so that some sorts of deformations, notably extensions, can increase the local random mixing of fluids with concentration gradients.  Shear flow, in the sense of parallel plate rheometer motion does not necessarily cause such mixing.  


The reason this should be troubling to those firmly convinced that an invariant based approach is sufficient is that, in classical rheology, the stress tensor is built on the deviator, $D_{ij}=\partial_{(i}v_{j)}$, and gives no unique decomposition of the flow into pure shear, extension and rotational parts.  The standard use of the term ``shear'' diffusion generally incorporates all nonrotational and translational changes in the local parcel.  Since we will show that such a decomposition is actually physically meaningful, we will refer to ``shear'' as implying the layered pure shear flow changes to make it distinct from extension for the elongational changes.  The symmetrization in the deviator will be shown to obliterate some essential physical information that would otherwise allow us to form such a decomposition that gives us the desired effects on mixing and a larger set of invariant variables to describe rheology.

The codeformational and corotational models \cite{Bird} follow a parcel in Lagrangian fashion and give forces based on the parcel's local deformation.  These models rely on the notion of ``objectivity'' which is the fluid dynamics analog of the relativistic principle whereby every observer sees the same physical laws.  The wrinkle here is that, while translation of a parcel is evident, the rotation of a deforming parcel is not.  On first consideration one might ask why the presence of a nonzero curl is not sufficient to demonstrate nonzero rotation of the local fluid parcels.  The deviator cannot distinguish between a pure shear and extension plus a dynamic rotation of the parcel.  Microscopically, the molecules in a pure shear can slide by in layers since the flow lines are linear and remain equidistant.  Extension requires rearrangements that must be irreversible.  This suggests that a rotating parcel undergoing pure shear and one undergoing extension exhibit different physical effects.  

Our investigation of deformation driven mixing will show that there is an absolute distinction between layered shear and extension and that local rearrangements at the microscopic level give a very real meaning to each.  This is one sense in which the flow contains missing invariants from approaches based on the deviator.  This allows us to construct a simpler picture of rheology where the local axes and rotation of the parcel has meaning defined by the derivatives of the unsymmetrized object $\partial_{i}v_{j}$.  One surprise will be that zero flow can sit at a nonanalytic point for some rheological features which dooms a general expansion approach in terms of tensors.  Specifically, the extension and compression of a parcel are shown to have different properties as far as mixing hence allow a similar possibility for rheological forces.  A similar situation arises in the theory of superconductivity and partially explains why it took fifty years to derive a theory after the effect was observed.  

The second part of this paper considers suspensions, specifically the kinds of flows that exist around particles suspended in shear flows.  It is shown that these give a kind of ``hidden energy'' that has to be included when seeking stationary  distributions of particles in flow gradients as happens in the Zweifach-Fung effect.  This stored energy must be accounted for in rheology where it can give a relaxation time not apparent from the bulk averaged velocity flow.  This is hardly surprising and is complimentary to the viscous corrections for suspensions done by Einstein \cite{Einstein} and others.  

The organization of the article will be as follows.  Invariants and the standard approach to rheology is briefly reviewed.  We then discuss mixing, in particular, the deformation driven mixing as opposed to thermal diffusion.  For the 2D case, we derive rules for unique flow decompositions based on maximizing the pure rotational component.  This gives a Fick's-type law for mixing with a causal correction.  The generalization to 3D is outlined and a theory of viscosity is given based on the flow decomposition.  
We conclude with a discussion on second order flow driven unmixing of fluids and thermal equilibrium of suspensions and concentration gradients in a stationary flow.  

\section{Standard Rheology}
Stokes laid out the foundations of hydrodynamics with a theory of parcels that are advected with the flow and driven by pressure and  stresses from internal dynamics and external forces that are now expressed in the Navier-Stokes equations (N-S)  
\begin{align}
\rho \left(\partial_{t}+v\cdot \nabla \right)v=-\nabla P+\nabla\cdot T+f
\end{align}
The stress tensor $T^{ij}$ is symmetric since the forces it applies to a parcel are from internal dissipative actions and so they cannot contribute angular momentum to it.  The velocity field can be expanded as $v_{i}=v_{i}^{(0)}+W_{ij}x^{j}+\ldots$.  To lowest order the stress tensor can only depend on $W_{ij}=\partial_{j}v_{i}$.  It is standard to construct quadratic (right or left Cauchy-Green) tensors from the symmetric traceless part of $W$, $D_{ij}=W_{(ij)}-\frac{2}{3}W^{k}_{k}\delta_{ij}$ as $D^{\dagger}D$ or $DD^{\dagger}$ and appeal to polar decomposition $D=RU$, where $U$ is a rotation at this point and argue that $U$ now is the part the does not contain rotation.  (This is true for the quadratic forms but, later, we will be interested in the linear form of $W$ itself and extract information from it directly with no quadratic mediator.)  

We are primarily interested in liquids where the density does not change.  Compressibility of a liquid is very low but spatial variation in density due to concentration or heating can be enough to cause measurable effects like convection over large enough scales.  Since this change in density leaves the particles in the same relative locations, the bulk viscosity seems not relevant.  It is possible we could have a solution with a heavier solute diffusing into a region to increase the local density.  This gives a net mass flow inwards.  The local number density of solvent and solute changes as well and it becomes unclear just what ``$v$'' is now supposed to mean.  The concentration change can alter the packing density of the parcel so it is not necessary that number density be preserved, however, these processes tend to be dominated by diffusion so that the flow itself is not losing energy to heat.  Bulk viscosity is small even in gases and it is not immediately obvious how a linear radial motion can dissipate at all.  
Internal energy changes associated with concentration imply thermodynamic variables will become important and generate forces as well.  Generally this is neglected since the effects are small.  Long term additivity of them can be important but we neglect them in our discussion here.  

Analyticity and symmetry imply the stress tensor must be made of the symmetrized functions of the derivatives of the velocity field.  This is usually built up in terms of the deviator, $D_{ij}=\partial_{(i}v_{j)}$, where the symmetrization eliminates dependence on any pure rotational motion.  Using the deviator, there are three nontrivial rotational invariants $I,II,III$.  These obey the Cayley-Hamilton theorem $D^{3}-I_{D} D^{2}+II_{D} D-III_{D} \text{I}=0$.  The most general function of these to give a stress tensor is built by scalar functions of the invariants $f(I,II,III)$ and the tensors available in the problem: $D_{ij}$, $\delta_{ij}$, $\epsilon_{ijk}$.  This approach was popularized by Rivlin in the 1950's \cite{Truesdell}.  Extensive work has been done using such models and they are standard in rheology \cite{Bird}.  Certainly, the use of microscopic strain is done explicitly incorporating relaxation time.  Ultimately, any liquid must be understood in such terms since, unlike gases where the momentum flux is entirely through lateral mass transfer, liquids transfer viscous forces through bond strain.\footnote{Superfluid Helium may be the one exception to this.}  The assertion here is that, microscopically, the motion of particles is dominated by diffusion with a slight bias due to flow and that such motions give an unambiguous local notion of rotation versus shear and extension.  We will next see though a discussion of mixing that symmetrization has removed more than just information about the pure rotation and give a process to  uniquely specify it.  

\section{Mixing} \label{Mixing}

For simplicity we consider a 2-D flow of a solution made of only two types of molecules with only isotopic changes in their composition.  The restriction to 2-D means there is only one shear and one extension mode for each parcel. 
The use of isotopes allows us to keep the geometric obstructions of each constituent to be the same while allowing the two kinds of molecules to be distinguishable.  

\begin{figure}
\begin{center}
\includegraphics[trim = 0mm 0mm 10mm 10mm, clip,width=6cm]{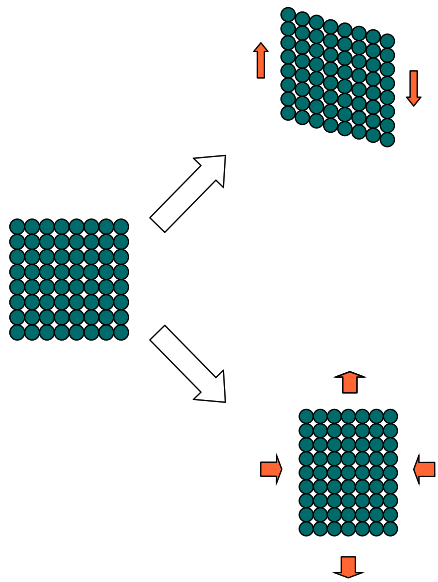}
\caption{A lattice structure under shear and extension deformations.}
\label{shearvext}
\end{center}
\end{figure}

We might try to model the liquid at each instant as a continuous random network.\footnote{It is possible that the fluid might form long lived clusters that effectively form microscopic solid blocks in the fluid.  This is a popular model for water but ultimately this only increases the granularity scale for mixing in what follows.}   Here we will use a simpler picture in terms of lattices.  This introduces an artificial order but our arguments will not depend strongly on it and all the pictures here could be replaced with more random ones with similar conclusions.  Our goal is a model where the specific details are not important so long as the thermal rearrangements are small compared to the ones induced by deformation from the flow.

The parcel deformations of shear and extension are shown in fig.\ \ref{shearvext}.  We see in the top right of the figure that the rearrangement due to shear flow is to translate one layer over another.  This may seem like a great simplification for a collection of particles that are vibrating rapidly and locally rearranging bonds (van der Waals or hydrogen) with their neighbors.  Evidence for its validity is that the thermal diffusion is low which implies that the geometric obstructions are enough to keep neighbours near each other despite these bond rearrangements.  This model also has some experimental verification from the case of a rotating Couette cylinder of soap solution with a drop of dye in it.  Many rotations will smear out the drop but unwinding the same number of times returns it very closely to its original state as a localized drop\cite{Batchelor}.  
This shows the rearrangements of shear flow are very nearly reversible.

\begin{figure}
\begin{center}
\includegraphics[trim = 30mm 0mm 100mm 120mm, clip,height=6cm]{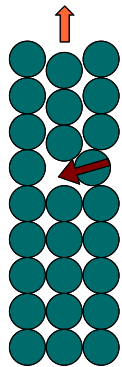}
\caption{A single particle rearrangement under extension deformation.}
\label{ext}
\end{center}
\end{figure}

By the reversibility of shear flow we consider its effect on the fluid fluid to be not mixing.  This means we can define the concentration as a scalar variable at each point that has been displaced but not otherwise altered by the velocity field.  This approximation should be valid until shear leads to gradients in concentration that are so steep that we cannot define any reasonably sized averaged parcel with microscopic homogeneity or the induced transverse rearrangements have built up over enough time (as in shear enhanced diffusion) that they can no longer be ignored.

\begin{figure}
\begin{center}
\includegraphics[trim = 10mm 20mm 10mm 0mm, clip, height=6cm]{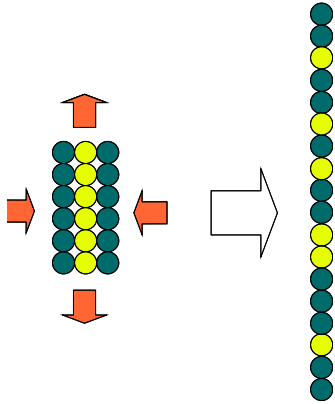}
\caption{Mixing rearrangements under extension deformation.}
\label{extmix}
\end{center}
\end{figure}

\begin{figure}
\begin{center}
\includegraphics[trim = 10mm 0mm 0mm 0mm, clip, height=6cm]{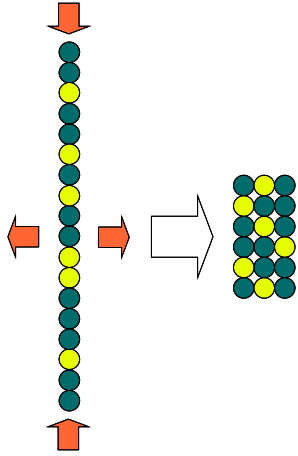}
\caption{The fluid parcel after the reversed extension deformation.}
\label{extmixrev}
\end{center}
\end{figure}

From this picture, (uniform) extension flow is seen to be different, fig.\ \ref{ext}.  
Extension has the effect of stretching some rows and eliminating others by drawing in molecules laterally towards the extension ``axis of elongation'' to fill the opening gaps. The process is random as to which neighboring molecules are chosen to fill them.  In the case of an isotopic gradient, such choices are destroying information about the concentration variation.  These choices are not reversible and so is reminiscent of thermal diffusion itself.  Considering the extreme case of three rows of fluid molecules in this simple lattice model, fig.\ \ref{extmix}, where we completely compress the three rows into one.  The choices of which color goes where is random and we end up in an irreversible mixed state.  Applying the reverse transformation the original order is lost, fig.\ \ref{extmixrev}.

The extreme result here is a result of the finiteness of the system and the extreme deformation we have applied to the fluid.  Any mixture that we could deform by extension down to a single layer thick will have all of its normal concentration gradient eliminated and be uniform as the extension is reversed.  More generally, if we make and N:1 change in the thickness of a parcel we can consider each set of N rows to get compressed into one.  
If the concentration gradient is small over N rows this will naturally be a very small effect.  Microfluidic systems are known for having serious mixing problems.  Here the granularity scale is not so many times less than the width of the system and one can repeat iterative actions many times rapidly.  The gradients in concentration will be generally softened by such actions as we see below.

\subsection{Deformation Mixing of Isotopic Solutions}

One could conceive of a number of higher order effects whereby molecules of different size and shape in solution have a bias in the probability of displacements in various flow fields and even a bias in migration rate for gradients of such flows.  To remove this complication let us consider a mixture of isotopes so they are electronically identical and only differ in nuclear mass.  This will affect phonon spectra, the inertia of the particles and diffusion rates but leave the deformationally driven mixing process minimally changed.  
To this end we specify a 2-D liquid of constant density $\rho$ and with velocity flow field $\vec{v}(\vec{x},t)$.  The fluid is made of 2 molecular types A and B (that only differ in the isotopes of the constituent atoms) with concentrations [molecules/Area]  $n_\mathrm{A}$ and $n_\mathrm{B}$ respectively.  We specify the flow field as a function of time and seek to find how the concentrations of A and B change in time.  Since $n_\mathrm{A} + n_\mathrm{B} = $constant we only need to solve for the equation of motion for $n_\mathrm{A}$.  

\begin{figure}%
    \centering
    \subfloat[Pure Shear]{{\includegraphics[trim = 0mm 0mm 0mm 0mm, clip, width=5cm]{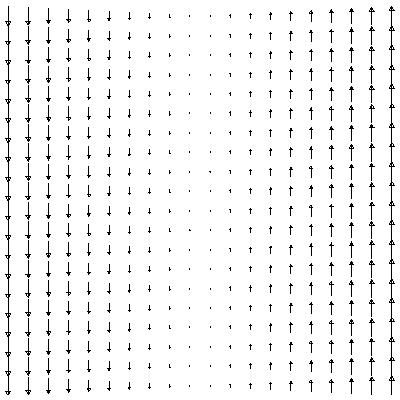} }}%
    \qquad
    \subfloat[Pure Extension]{{\includegraphics[trim = 0mm 0mm 0mm 0mm, clip, width=5cm]{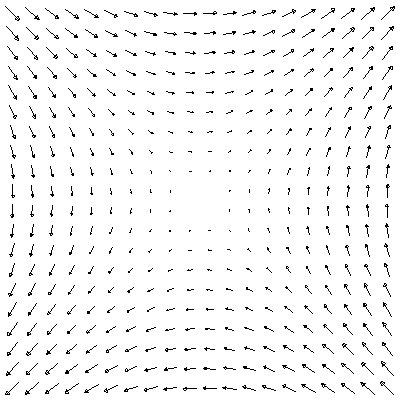} }}%
    \caption{Pure shear vs. extension flow.}%
    \label{fig:flows}%
\end{figure}

The deviator $D_{ij}$ is the same for the flow fields $v=2\alpha y\hat{x}$ and $u=\alpha y\hat{x}+\alpha x \hat{y}$.  The difference in these is illustrated in fig.\ \ref{fig:flows}.  
When we evaluate $\nabla\cdot T=\nabla\cdot (\eta D)$, one of the two terms vanishes because $\nabla\cdot v=0$ and the correct viscous damping force direction appears in the N-S equations.  For simple fluids this gives us no problem.  However, for the case of mixing, these two flows are quite different and we need the full unsymmetrized $\partial_{i}v_{j}$ to distinguish them.  

If we assume a general local parcel that is undergoing extension, shear and rotation we can write
\begin{align}\partial_{i}v_{j} &=\Omega+E+S=  \left(\begin{array}{ccc} 
      \epsilon & r \\
      -r & -\epsilon \\
   \end{array}\right)+
   R(\phi)SR^{\dagger}(\phi)
   \end{align}
   where the shear matrix $S$ is chosen as
   \begin{align}
   S_{x}=\left(\begin{array}{ccc} 
       0& \alpha \\
       0&0  \\
   \end{array}\right) \text{~~~~~or~~~~~~}  S_{y}=\left(\begin{array}{ccc} 
       0& 0 \\
       \alpha&0  \\
   \end{array}\right)
\end{align}
In these coordinates, the extension is along the x and y axes and the shear parallel to the $x'=R(\phi)x$ axis or $y'=R(\phi)y$ respectively.  We can see that there are four parameters $(\epsilon, r, \phi, \alpha)$ that specify this matrix, however, the constraint Tr$(\partial_{i}v_{j})=0$ tells us that this is overdetermined.  Furthermore, we could view this in any rotated coordinate system where the result is not so simple.  
The most convenient coordinate systems are the ones where shear or extension is along the coordinate axes.  This arises from microscopic considerations and we expect $\phi=\phi(\epsilon, r, \alpha)$.  Some larger scale considerations may give natural choices.  For example, given a function that specifies local heating from a given shear and extension, nature may choose the one that has the least energy loss.  In the absence of such a minima, nature might choose to minimize some thermodynamic quantity or the shear direction may wander in such a regime randomly in response to hidden microscopic variations.    Alternatively, nature may choose for us to maximize the rotation in the decomposition.  Since this part of the deformation cannot contribute forces it is a likely choice and the one we will use in the following sections.  For now we assume such a decomposition is given and derive the mixing equation from deformational flow.  

The flow now gives a natural local coordinate decomposition, which we call the flow intrinsic decomposition, $(\hat{e}_{1},\hat{e}_{2})$.  If extension vanishes we can use the shear direction and its normal: $(\hat{s},\hat{s}^{\perp})$ but, since mixing is only a function of extension, will not matter for this discussion.  
Let the solution concentration of component B be given by $\mathpzc{c}(x)=n_{B}(x)$.  The diffusion current, in the rest frame of the parcel, is only a function of the extent of extension occurring in the inwards direction and the curvature of the concentration along this axis.  
\begin{align}
j={D}  \cdot(\nabla_{e_{i}}\nabla_{e_{i}}\mathpzc{c}\cdot \nabla_{e_{i}}v_{e_{i}})\hat{e}_{i}
\end{align}
where $\hat{e}_{i}$ is the extension direction under which the flow is contracting i.e.\ the $\hat{e}_{i}$ that gives Min$(\nabla_{e_{1}}v_{e_{1}},\nabla_{e_{2}}v_{e_{2}})$ so that $\nabla_{e_{i}}v_{e_{i}}<0$.  The concentration evolves as
\begin{align}\label{ConvDiff}
\dot{\mathpzc{c}}+\nabla\cdot(\mathpzc{c}v)=\nabla\cdot j
\end{align}

Conservation is automatic and the concentration of the ``solute'' obeys $\dot{n_{A}}=-\dot{\mathpzc{c}}$.  Such ``slow diffusion'' equations do have causality problems.  Using that the intrinsic velocity scale for the medium is the speed of sound, $c_{s}$, we can find the maximum local velocity of propagation, relative to the flow field, as $v_{p}=j/c_{s}$.  Allowing the diffusion constant $D$ to become a function we can enforce that $v_{p}\le c_{s}$ while obeying Fick's law at the low current regions that dominate realistic experiments.  The only length scale in the problem, $d$, is the granularity scale of the liquid.  The diffusion constant, $D$, has units $[l^{3}]$ so $D\sim d^{3}$.  

If the liquid is a complex fluid then we expect shear and extension to give different results.  Shear thickening and extension thickening are common in polymer melts.  This is generally modeled by the use of invariants functions designed to give these limits with some nonlinear smooth transition between them.  The rheology in terms of such a flow decomposition is simple if the shear and extension flow contributions decompose to give linearly independent forces.  By using the intrinsic flow decomposition we can specify the induced stresses in terms of the deformations.
\begin{align}\label{stress}
T=\eta_{e}(\nabla_{e_{i}}v_{e_{i}})\hat{e}_{i}\otimes\hat{e}_{i}+\eta_{e}(\nabla_{e_{o}}v_{e_{o}})\hat{e}_{o}\otimes\hat{e}_{o}+\eta_{s}(\nabla_{e_{s}^{\perp}}v_{e_{s}})\hat{e}_{s}\otimes\hat{e}_{s}
\end{align}
The notation $\nabla_{e_{i}}v_{e_{i}}$ here indicates the gradient of the velocity component along $\hat{e_{i}}$ projected along $\hat{e_{i}}$.  The sign of $e_{s}^{\perp}$ is ambiguous here.  It should be chosen so the effect is to damp the motion.  Naturally we expect to choose it so we obtain positive $\eta_{s}$.  The expression is manifestly symmetric.   Shear flow must induce a counter torque so that no internal rotation is generated as a function of deformation.  

At this point we urge the casual reader to carefully consider the arguments leading to this.  Generally such an expression is considered inadmissible based on the usual invariance paradigm of construction.  The presence of microscopically determined and unique shear and extension axes renders it physically meaningful in the same way that dielectric response has invariance from the meaningfulness of the frame of the dielectric, even if it is much uglier in a boosted frame.  Indeed, the whole point of this article is to argue that the invariant construction methods are lacking and will never be adequate to describe general rheology and provide a direct path between the material parameters and the microscopic dynamics.  
We will see that there are natural reasons the decomposition into these variables need not be everywhere smooth even for globally smooth flows.  

\section{Minimal Decompositions} 
Here we seek to find unique physically motivated way to derive a decomposition of a flow into rotation, extension and shear components: $\partial_{i}v_{j}=\Omega+E+S$.  

\begin{align}W=\Omega+E+S&=  \left(\begin{array}{ccc} 
      \epsilon-s\cos(\phi)\sin(\phi) & r+s\cos^{2}(\phi) \\
      -r-s\sin^{2}(\phi) & -\epsilon+s\cos(\phi)\sin(\phi) \\
   \end{array}\right)\\
   &= \left(\begin{array}{ccc} 
      \epsilon-\frac{1}{2}s\sin(2\phi) & r+\frac{1}{2}s(\cos(2\phi)+1) \\
      -r+\frac{1}{2}s(\cos(2\phi)-1) & -\epsilon+\frac{1}{2}s\sin(2\phi) \\
   \end{array}\right)\\
 &=\left(\begin{array}{ccc} 
      \epsilon & r+\frac{1}{2}s \\
      -r -\frac{1}{2}s& -\epsilon\\
   \end{array}\right)
   +\frac{1}{2}s \left(\begin{array}{ccc} 
      -\sin(2\phi) & \cos(2\phi) \\
      \cos(2\phi) & +\sin(2\phi) \\
   \end{array}\right)
\end{align}

This is based on a particular coordinate system and our local data is of the form
\begin{align}H=\partial_{i}v_{j} =  \left(\begin{array}{ccc} 
      a & b\\
      c & -a \\
   \end{array}\right)\\
\end{align}
so that our decomposition is underdetermined.  Let us take the point of view that, since rotation of a parcel is dissipationless, nature chooses to maximize the rotation $r$ of the flow field.  Solving for the variables $(r,s,\epsilon)$ in $W=H$ we have 
\begin{align}
r=-\frac{c \cos^{2}(\phi)+b \sin^{2}(\phi)}{ \cos^{2}(\phi)-\sin^{2}(\phi)}
\end{align}
The extrema for this occurs at $\phi=0,\frac{\pi}{2},\pi,\frac{3\pi}{2}$ with maxima determined by the signs of $c$ and $b$.  Furthermore, if we rotate the matrix $W$ by and angle $\theta$ and solve $R(\theta)WR^{\dagger}(\theta)=H$ we find that the maximum rotation $r$ is found at one of the points
\begin{align}\label{angle}
\theta=\pm\arctan \left( \frac{1}{2}\,{\frac {\pm(b+c)+\sqrt {4\,{a}^{2}+(b+c)^{2}}}{a}} \right)
\end{align}  
and the maximal angle of shear relative to this is $\phi=0$ or $\phi=\pi/2$.  The negative sign is generally the physical one but the positive one is retained as a possibility since a nonanalytic point arises as the extension vanishes.   If we choose coordinates so the outwards extension is along the x-axis then $a>0$ and so the rotation is counterclockwise then $r>0$.  In these rotated (and possibly inverted) coordinates the rotation decomposition is of the form
\begin{align}W=  \left(\begin{array}{ccc} 
      \epsilon & r+s \\
      -r & -\epsilon \\
   \end{array}\right)=H'=  \left(\begin{array}{ccc} 
      a' & b' \\
      c' & -a \\
   \end{array}\right)
\end{align}
from which we can read of $\epsilon=a',r=-c',s=b'+c'$.  The stress in eqn.\ \ref{stress} is determined by $\hat{e}_{o}=P\cdot R(\theta)\hat{x},\hat{e}_{i}=P\cdot R(\theta)\hat{y},\hat{e}_{s}=P\cdot R(\theta+\phi)\hat{x}$ where $P=\pm1$ is the parity of the coordinate transformation so that rotation is ccw.  

Alternate approaches could be to assume that nature chooses to minimize the total damping or maximizes shear over extension flow.  A microscopic description would seek to specify the extent of rearrangements through sliding of layers versus forced random rearrangements of particles.  Extended and deformable particles will have orientational and extensional features also dependent on the smoothed flow.  The global velocity flow is actually a coarse grained smoothed function of these granular motions yet we presume this is sufficient determine the microscopic motions statistically.  Based on our analysis, it is generally sufficient to specify the local orientation of the extension angle relative to the coordinates as $\theta(a,b,c)$.  Once this is known, the shear is parallel to $\hat{e}_{i}$ or $\hat{e}_{o}$, and the extent of the extension, the rotation and the magnitude and sign of the shear are all determined from $W(x)$.  Since there can be nonanalytic features to the angle function $\theta$, as in eqn.\ \ref{angle}, we anticipate that smooth flows of complex fluids may have nonanalytic points due to qualitative microscopic changes in the nature of the rearrangements.  

Higher order considerations are difficult in the invariant approach to rheology.  Higher derivatives do not give tensors but live in jet spaces with transformation rules that are affine.  Generalization in this formalism is potentially more straightforwards.  The bias in physics towards tensor treatments is due to an ease of enforcing invariant properties and an expectation of analytic expansions.  Here we have presumed that microscopic rearrangements are the fundamental feature to retain rather than striving for the simplest or most familiar mathematics.  For higher derivatives one already has a preferred decomposition in terms of linear flows.  In the linear case, we gave a method to derive the intrinsic rotation, $r$, of a parcel from its coarse grained flow.  Higher order flows may induce other characteristic directions that one can use to build response in terms of.  Later we will discuss how higher order flows can inherit granular biases in transport that can suggest further such intrinsic geometry.  

\section{3D Flows}
The extension of this theory to 3D has some additional complications.  Firstly, is the number of degrees of freedom (DOF). The linear data for incompressible flow, $\partial_{i}v_{i}$, has eight DOF.  The symmetrization of it, the deviator $D_{(ij)}$, has only five so is even more deficient than in the 2D case.  Extension can now have either one or two inwards moving axes that drive deformational diffusion.  This set of axes requires three Euler angles, $(\theta_{e},\phi_{e},\psi_{e})$, to orient it and two more to specify the strength of the flows, $(\epsilon_{1}, \epsilon_{2})$.  Shear requires two angles to orient, $(\theta_{s},\phi_{s})$, and one for strength, $s$.  Similarly, rotation requires three parameters, $(\theta_{r},\phi_{r},r)$.  This is similar to the 2D case where we had to specify the five intrinsic parameters $(r,s,\epsilon,\phi,\theta)$ using the three flow parameters $(a,b,c)$ except now we have eleven variables and eight constraints.  By maximizing $|r|$, or using some other dominant physical consideration, we can find the optimal rotation axis, $(\theta_{r},\phi_{r})$, and intensity, $r$, and determine the remainder of the functions.  

In the case of linearly decomposable forces, the rheological stress induced by the flow will be a function of the three extension axes $\hat{e}_{i}=R(\theta_{e},\phi_{e},\psi_{e})_{ij}\hat{x}_{j}$.  
\begin{align}
T=\eta_{e}\bigg((\nabla_{e_{1}}v_{e_{1}})\hat{e}_{1}\otimes\hat{e}_{1}+(\nabla_{e_{2}}v_{e_{2}})\hat{e}_{2}\otimes\hat{e}_{2}+(\nabla_{e_{3}}v_{e_{3}})\hat{e}_{3}\otimes\hat{e}_{3}\bigg)+\eta_{s}(\nabla_{e_{s}^{\perp}}v_{e_{s}})\hat{e}_{s}\otimes\hat{e}_{s}
\end{align}

Mixing driven diffusion is more complicated because we only expect the inwards extension flow and their projection parallel to the concentration gradient to contribute.  The diffusion current is then given by
\begin{align}
j={D}  \cdot\left(\sum_{i}(\nabla_{e_{i}}\nabla_{e_{i}}\mathpzc{c})\cdot \nabla_{e_{i}}v_{e_{i}}\right)\hat{e}_{i}
\end{align}
where the sum is over the indices for which $\nabla_{e_{i}}v_{e_{i}}<0$.  (Here we have an unfortunate violation of the Einstein summation convention.)  The sum is over at most two terms since at least one of the extension directions must have an outwards flux.  

\section{Gradient Flows of Solutions and Suspensions}

We have seen how extension flow forces mixing.  We can give a minimum diffusive drift due to extension: For an N:1 change the nearest particles will drift $\sim$N/2 places from them.  The concentration changes are only driven by the \textit{inwards} component of the extension flows.  Details of curvature in the concentration tend to get obliterated in the bulk and linear concentrations are smoothed out at boundaries (as we can see by considering the boundary to have a reflected concentration at the wall so the edge becomes a point of divergent curvature).  Here we will look at how gradients in the shear or extension flow combined with a bias in the hopping of solute versus solvent particles or hydrodynamic force biases on suspended particles can produce ``unmixing'' in the form of equilibrium concentration gradients.  We will investigate analogs to Perrin's diffusion gradient experiment \cite{Perrin}  where the thermal diffusion is biased by hydrodynamic forces in a driven nonequilibrium flow instead of gravitational ones of an equilibrium suspension.  The Zweifach-Fung \cite{Zweifach}\cite{Fung} effect is an example of such a sorting mechanism in blood flow that seems closely related to this problem.  

\subsection{Hidden Local Flows and Thermodynamics} 

We are generally interested in solutions and suspension where the flow variations from the inclusion particles gives changes over a volume that are small compared to the scale of change of the macroscopic flow.  In micro and nanofluidics this may not be always true where extreme shears are possible.  The low density limit is implicit so that the flow variations around such particles do not overlap.  We will compute results for such small flow cases but should remember we could consider such effects with higher order effects even at low densities of impurities.  

Einstein \cite{Reif} gave the result for such a suspension of spherical inclusions in a fluid with density $\rho$ and viscosity $\eta$.  The hydrodynamics obey the N-S equations with modified density $\rho'\approx \rho+(\rho_{s}-\rho)\eta$ and $\eta'=\eta(1+5\nu/2)$, where $\nu$ is the volume fraction of inclusions.  This is due to the intensified flow and steeper local flow gradients strongest very close the included particles.  We can understand this from the fact that the shear flow about the inclusions is increased by $\delta v \sim d\nabla v$ so that $T$ changes by $\sim \eta d \nabla^{3}v$ over a volume $\sim d^{3}$ so the additional stress per volume is $\delta T \sim \eta n d^{4}  \nabla^{3} v=\eta n d^{3}\nabla^{2}(d\nabla v)$ however this is not the same $v$ as if the macroscopic flow had no inclusions.  Furthermore, we don't want to reference the microscopic flow in our final result.  
The inclusions and their surrounding flow modifications create a kind of ``excluded volume'' so that the net coarse grained $\nabla v_{macro}$ is steeper and the hidden local flow losses are compensated by a larger effective viscosity for the the macroscopic flow.  The stress tensor $T=\eta\nabla v$ has $\nabla_{i} v_{j}\rightarrow \nabla_{i} v_{j}^{macro}+\omega_{ij}$.  The net viscous force directions are not changed by inclusion so $f=\nabla\cdot T=\eta\nabla\cdot \nabla v_{macro}+\eta\nabla\cdot\omega=(\eta+\delta\eta)\nabla\cdot \nabla v_{macro}$ where $\delta\eta\sim \eta n d^{3}$.  (Unless otherwise specified, from here on, $v$ will refer to the macroscopically smoothed velocity field.)  

We now are in a position to consider an effect that is often missed in this averaging.  The viscosity increases because of the enhanced small scale flow about the inclusions but we also have an increase in flow kinetic energy that is above what we derive based on the averaged density and regional averaged flow: $\rho'v^{2}_{macro}$.  For pure shear, there are lateral motions and enhanced velocity near the edges as the flow accelerates around the sides.  As we bring up the flow to speed we have to input energy for the net kinetic motion, damping \textit{and} this ``hidden'' kinetic energy.  We can estimate this extra energy by the velocity increase in the near field flow around the inclusion.  $v\approx v_{mean}\pm d\nabla v$.  The resulting increase in the kinetic energy density is $\nu (d^{3}\rho) G d^{2} (\nabla v)^{2}$ where $G$ is a geometric factor of $\mathcal{O}(1)$.  Now let us apply this to the case of Poiseuille flow down a long cylindrical pipe.  The solution is \cite{Batchelor}
\begin{align}
v_{z}=\frac{1}{4\mu}\frac{dP}{dz}(r^{2}-R^{2})=v_{0}\frac{R^{2}-r^{2}}{R^{2}}
\end{align}
where $v_{0}$ is the flow velocity at the center of the pipe.  The hidden kinetic energy change per particle is $\mathcal{K}=4\rho G d^{5} v_{0}^{2}r^{2}/R^{4}$.  The equilibrium distribution of particles is analogous to the barometric distribution 
\begin{align}
\nu(r)=\nu_{0}\exp\left(-\frac{\mathcal{K}(r)}{k_{B}T}\right)
\end{align}
  For very strong shears $\sim1$m/s in micron sized channels this can give a strong bias of suspended particles away from the wall (independent of the volume exclusion effect that acts within one particle radii of the walls).  In the case of an uneven splitting of the flow at a junction, as in the Zweifach-Fung effect, the faster flow channel gets a disproportionate fraction of the particles.  This has been suggested as due to a boundary exclusion effect \cite{Doyeux} but the role of the thermodynamic minima of such particles in a flow can now be considered as well.  For example, in such uneven flows, is the minimum channel in the slower channel even connected to that in the input channel?  If not, suspended particles would require a thermal fluctuation to cross the barrier between them.  

The details of the hidden KE of the flow are only approximately quantitative here.  If one considers a 2D pure extension flow about a particle, one sees that there is much less need to distort the flow so the that the hidden energy should also be a function of the flow type near it.  This also suggests that, even for a simple fluid, the suspension will take on complex rheological properties even when the suspended particles and their flow fields remain widely separated.  The overlapping of the flow fields around the particles  give nonlinear terms that favor separation.  The shear flow over a particle will include some local extension so induces some mixing of microscopic solutes there.  

We now can ask what happens in the limit that the suspended particles get arbitrarily small.  Furthermore, let them be made of the same material as the suspended fluid, perhaps at the freezing point where the small clusters are in the solid phase.  This leads to an interesting paradox.  As we diminish the size of the particles while increasing their number so the total volume fraction remains the same, Einstein's relation predicts the same bulk averaged viscosity.  However, this is just the limit of the pure fluid which must give the viscosity of the pure solute.  This suggests that there is some scale at which flow driven rearrangements cease to be important.  

One reason this is interesting is that we care about dissolved two component solutions where the model of a suspension as a mixture of granular particles in a hydrodynamic background is no longer valid.  The above argument should have some meaning for solutions especially when one set of molecules or ions has a solvation size much larger than the other.  In the case of a solution where the viscosity is unchanged as a function of concentration but the density of the two components are different one can enquire about the difference in kinetic energy of the flow based on radial mass redistribution due to concentration gradients.  Higher mass particles will diffused to the center but experience an energy cost in joining the faster flow.  Unlike the above example, this extra energy is not locally determined but is frame dependent.  It also has implications for the pressure of the driving field as these transitions occur.  For this reason, it is not immediately clear that laminar flow will create a bias of the higher weight component near the walls and how we should attack the problem of the equilibrium of this particular driven flow.  Since the thermal velocity of light molecules is typically hundreds of m/s we expect such an effect to be very small compared to the shearing velocities that are below the turbulent Reynolds numbers.  Nevertheless, iterative processes might generate larger separations and the effect could be larger for heavier constituents.  

\subsection{N-S with ``Hidden'' KE}
In the above discussion we have considered the viscous and mixing effects of suspensions but have neglected that the N-S equations have to provide this hidden energy from forces.  The N-S equations are essentially momentum conservation laws.  As wall forces or pressure gradients create a flow pattern starting from rest, there is additional damping due to the local shears at the particles and the enhanced shear they induce between them but the $\frac{1}{2}\rho v^{2}$ kinetic energy that it has created (where $v=v_{macro}$) is not all the kinetic energy present.  This energy must be accounted for in the description.  To this end we introduce a hidden reservoir of stress and energy encoded in the form of a local variable $\kappa_{ij}$ that tends to relax to a function of the local flow $W_{ij}=\partial_{i}v_{j}$.  It is an advected quantity that will be a source and sink of stress for the observable macroscopic flow especially in the case of strongly rapidly and varying flows.  

For a given macroscopic flow, in a parcels rest frame, we know there is a hidden microscopic flow about the suspended particles.  When this parcel has been subjected to this flow steadily for a while we expect $\kappa$ to converge to $\kappa_{0}(W)$, a function of the local flow profile.  This suggests an equation of motion for $\sigma$ of the (comoving) form
\begin{align}
\dot{\kappa}=-\gamma (\kappa-\kappa_{0}(W))
\end{align}
The N-S equations are modified by a the back reaction transient stress term $\kappa$ so that the final equations of motion are
\begin{align}
\rho \left(\partial_{t}+v\cdot \nabla \right)v&=-\nabla P+\nabla\cdot T-\gamma^{-1}\nabla\cdot\dot{\kappa}\\
\dot{\kappa}+v\cdot\nabla \kappa&=-\gamma (\kappa-\kappa_{0}(W))
\end{align} 
This has the potential to create significant lags in response when the hidden kinetic energy of the flow is large compared to the local change in the kinetic energy on the scale of $\Delta t=\gamma^{-1}$.

\subsection{Extension Gradients}
The special distinction of extension flow versus shear flow motivated our discussions on mixing and then rheology.  This has all been in respect to linear aspects of the flow.  The inwards directions of extension flow have the capacity to smooth out and obliterate concentration information.  This begs the question of how gradients of extension behave.  In particular we would like to study the possibility of flow itself driving migration of one fluid constituent relative to the others.

\begin{figure}
\begin{center}
\includegraphics[trim = 0mm 140mm 0mm 0mm, clip,height=3cm]{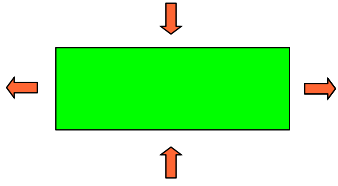}
\caption{Uniform extension flow.}
\label{extblock}
\end{center}
\end{figure}

\begin{figure}
\begin{center}
\includegraphics[trim = 0mm 70mm 0mm 0mm, clip,height=4cm]{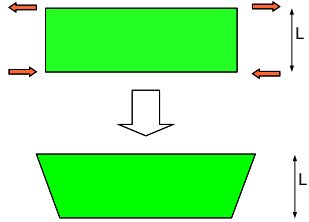}
\caption{Zero net shear and extension with an extension gradient in the vertical direction.}
\label{puregradext}
\end{center}
\end{figure}

\begin{figure}
\begin{center}
\includegraphics[trim = 0mm 60mm 0mm 80mm, clip,height=4cm]{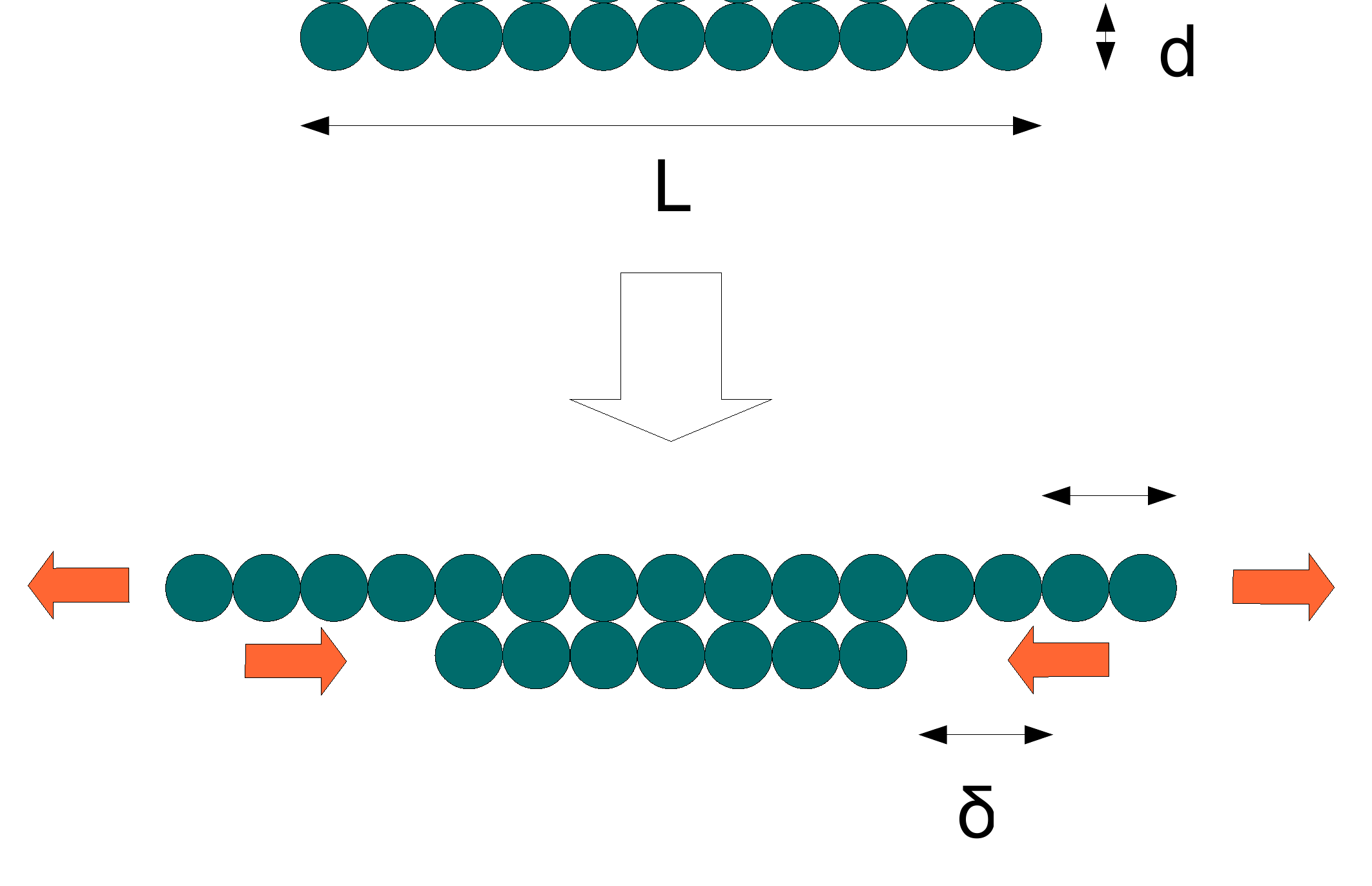}
\caption{Two layers of a parcel on either side of the center of mass in the gradient of an extension flow.}
\label{layers}
\end{center}
\end{figure}

Consider the simple 2D extension flow in fig.\ \ref{extblock} compared to that in fig.\ \ref{puregradext}.  The second flow has a net displacement of mass as particles are shifted to upper layers as in fig.\ \ref{layers}.  
This flow is described by
\begin{equation}
\vec{v} = \alpha\, x y\,\hat{x}
\end{equation}
In fig.\ \ref{layers}, we see two layers of molecules on either side of the center of mass of the parcel.  The axis of increasing extension is oriented in the y-direction.  The molecules are represented as circles of diameter d and the number of molecules per layer is $N = Ld/(d^2) = L/d$.  Let the liquid be composed of two equally sized but chemically dissimilar components A and B (as distinct from the isotopic case in Sec.\ \ref{Mixing}).  The action of shearing drives a flux of $\alpha L/d$ particles upwards.  

We now are interested in a bias in the dislocation choices driven by the deformation of particles of type A and B.  It is possible that one type is heavier or has a stronger binding to its neighbors. If there is no difference in A and B then this will be proportional to the number density of each particle type.  A reasonable hypothesis is that the bias, $\beta$, is also function of the relative diffusion rates of the the particle types in the solution.  The simplest case being
\begin{align}
\beta_{A}=\frac{D_{A}n_{A}}{D_{A}n_{A}+D_{B}n_{B}}\\
\beta_{B}=\frac{D_{B}n_{B}}{D_{A}n_{A}+D_{B}n_{B}}
\end{align}
The current along the vertical axis of extension gradient, $\hat{e}_{g}$ is then 
\begin{align}
j_{A}=\beta_{A} (\nabla_{\hat{e}_{g}}v^{\perp})\hat{e}_{g}\\
j_{B}=\beta_{B} (\nabla_{\hat{e}_{g}}v^{\perp})\hat{e}_{g}
\end{align}
where $v^{\perp}$ is the perpendicular flow to $\hat{e}_{g}$.  The concentration then evolves by the convective diffusion equation in eqn.\ \ref{ConvDiff}.  This is clearly a small effect.  A unit extension gives a net advance of one constituent over the other by, at most, one molecular layer.  Nanofluidic systems can be a relatively small number (hundreds or thousands) of molecular diameters across and a simple deformation action can be repeated many times.  This suggests a way to separate constituents by the iteration of extensions at least for such small systems.  

\section{Conclusions}
The granular nature of liquids has been the motivation for our discussion of mixing, rheology and some exotic nonequilibrium thermodynamic forces in complex fluids, suspensions and solutions.  While this discussion can hardly be considered a complete analysis, it suggests a broader range of dynamics with potential practical applications for confined fluids under extreme deformations.  Given the broad range of problems with such fluids in the chemical and medical industries, these investigations have much more than purely theoretical value.  The mathematical foundations of rheology are elegant and long established yet they have not yielded the kind of success in modeling polymer melts and other complex fluids that was initially hoped.  Additionally, there are numerous exciting problems seeking resolutions in fluid mechanics including the effects of very dilute polymer solutions in reducing turbulent drag, hydrophobic and dimpled surfaces, and non-Ahrreius viscous behavior.  Having a fully general framework for a rheology provides better opportunities to match with microscopic models and, presumably, fewer necessary parameters to model it generally.  

We have suggested ways the granularity scale of a liquid can lead to unmixing of constituents with particular flows and given a measure of the ``hidden'' kinetic energy of flows in suspensions with associated lag times in the response.  This analysis has only been to low order and not investigated deeply the possibilities for nonanalytic behavior or how relaxation can give ``kinks'' and abrupt topological jamming in constituent molecules as a result of these finite lag times.  Each of these seem deserving of significantly more investigation.  Although the author is critical of the ``math motivated'' approaches to fluid dynamics and rheology, it may be that there may yet be some great synthesis from the theory of jet spaces provided sufficient physical reality is encoded in the continuum variables that describe the fluid.  From the simple examples here, even second order effects introduce surprising complications.

\end{document}